\documentclass[preprint,aps]{revtex4-1} 

\usepackage{epsf}
\usepackage{epsfig}
\usepackage{color}
\usepackage{graphicx}
\usepackage{rotating}

\begin{document}

\title{The Interplay Between Lattice Topology, Frustration, and Spin Quantum Number in Quantum Antiferromagnets on Archimedean Lattices}

\author{D. J. J. Farnell}

\affiliation{School of Dentistry, Cardiff University,  Cardiff CF14 4XY, Wales, United Kingdom}

\author
{
O. G\"otze, J. Schulenburg,  and R. Zinke 
}
\affiliation
{
Institut f\"ur Theoretische Physik, Universit\"at Magdeburg, D-39016 Magdeburg, Germany
}

\author
{
R. F. Bishop and P. H. Y. Li
}
\affiliation
{
School of Physics and Astronomy, The University of Manchester, Schuster Building, Manchester M13 9PL, United Kingdom
}

\date{\today}

\begin{abstract}
\noindent 
The interplay between lattice topology, frustration, and spin quantum number, $s$, is explored for the 
Heisenberg antiferromagnet (HAFM) on the eleven two-dimensional Archimedean lattices (square, honeycomb, 
CaVO, SHD, SrCuBO, triangle, bounce, trellis, maple-leaf, star, and kagome). We show the CCM provides 
consistently accurate results when compared to the results of other approximate methods.
The CCM also provides valuable information relating to 
the selection of ground states and we find that this depends on spin quantum number for the 
kagome and star lattices. Specifically, the $\sqrt{3}\times\sqrt{3}$ model state provides lower 
ground-state energies than those of the $q=0$ model state for the kagome and star lattices for 
most values of $s$. The $q=0$ model state provides lower ground-state energies only for 
$s=1/2$ for the kagome lattice and $s=1/2$ and $s=1$ for the star lattice.
The kagome and star lattices demonstrate the least amount of magnetic ordering and 
the unfrustrated lattices (square, honeycomb, SHD, and CaVO) demonstrate the most magnetic 
ordering for all values of $s$. The SrCuBO and triangular lattices also demonstrate high 
levels of magnetic ordering, while the remaining lattices (bounce, maple-leaf, and trellis) 
tend to lie between these extremes, again for all values of $s$. 
These results also clearly reflect the strong increase in magnetic order with 
increasing spin quantum number $s$ for all lattices. The ground-state 
energy, $E_g/(NJs^2)$, scales with $s^{-1}$ to first order, as expected from 
spin-wave theory, although the order parameter, $M/s$, scales with $s^{-1}$ 
for most of the lattices only. Self-consistent spin-wave theory calculations indicated 
previously that $M/s$ scales with $s^{-2/3}$ for the kagome lattice HAFM, whereas previous 
CCM results (replicated here also) suggested that $M/s$ scales with $s^{-1/2}$. It
is probable therefore that different scaling for $M/s$ than with $s^{-1}$ does indeed occur 
for the kagome lattice. By using similar arguments, we find here also that $M/s$ scales 
with $s^{-1/3}$ on the star lattice and with $s^{-2/3}$ on the  SrCuBO lattice. \\ \\
\noindent 
{\bf Keywords:} Archimedean Lattices; Heisenberg Antiferromagnets; High-Order Coupled Cluster Method (CCM) 
\end{abstract}

\maketitle
\section{Introduction}

\begin{figure}
\begin{center}
\epsfxsize=13cm
\centerline{\epsffile{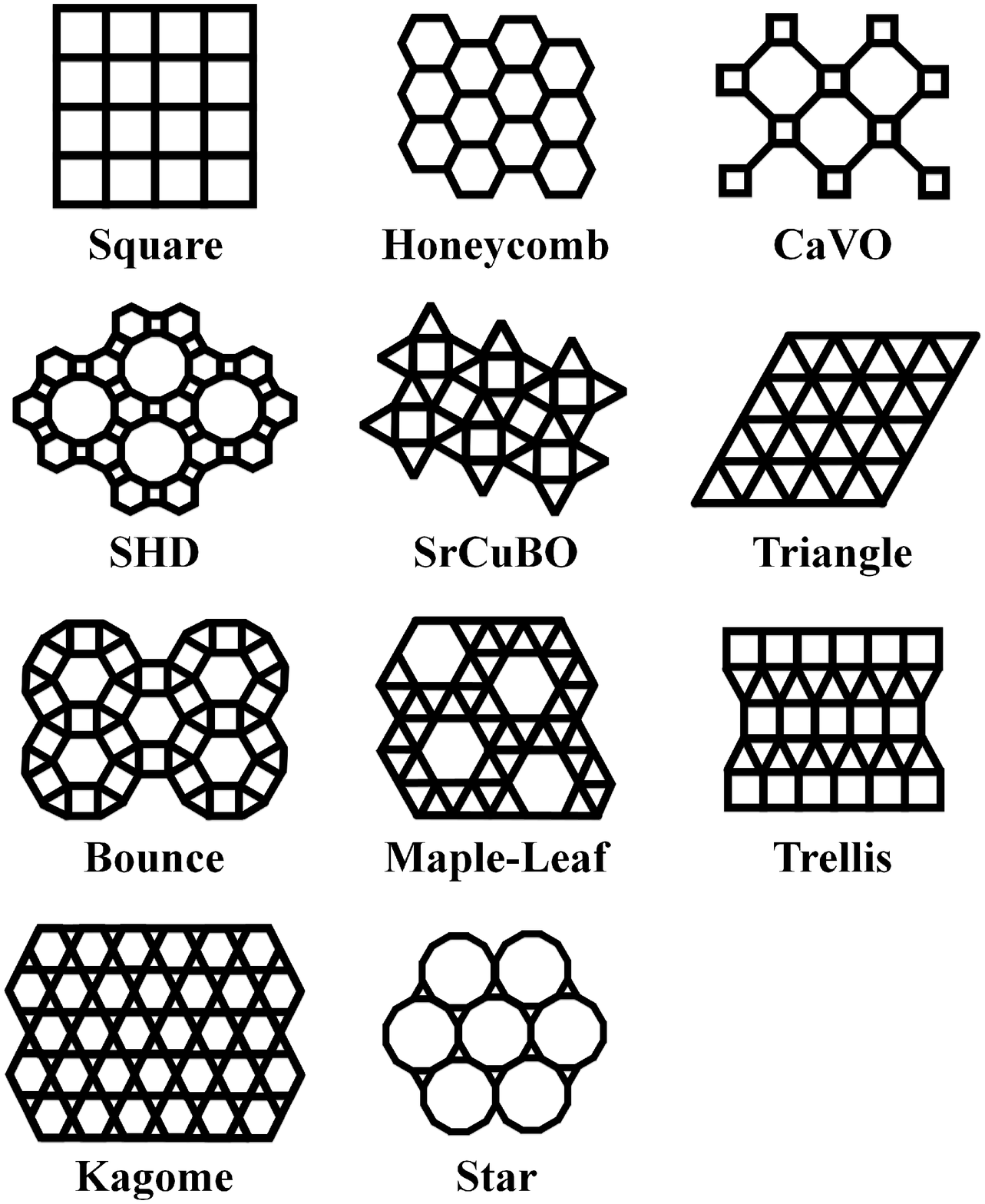}}
\caption{\label{fig1}The eleven Archimedean lattices: square (4$^4$), honeycomb (6$^3$), CaVO (4.8$^2$), 
SHD (4.6.12), SrCuBO (3$^2.4.3.4$), triangle (3$^6$), bounce (3.4.6.4), trellis 
(3$^3$.4$^2$), maple-leaf (3$^4$.$6$), star (3.12$^2$), and kagome (3.6.3.6). 
The mathematical description (in brackets) given by numbers $n_i$ 
separated by dots (i.e., $n_1 . n_2 \cdots n_r$) corresponds to the number of vertices of the polygons arranged 
around a vertex for each lattice.}
\end{center}
\end{figure}

Archimedes was one of the first people to describe regular tilings in two spatial 
dimensions. Archimedean lattices  \cite{ccm_high_order_archimedean_2014,richter2004a,unjong2015,zheng2014}
are infinite and they are composed of two-dimensional arrangements of regular polygons 
with every site equivalent. 
As shown in Fig. \ref{fig1}, there are eleven uniform two-dimensional Archimedean lattices. 
Such Archimedean ``bodies'' were gradually rediscovered in 
the Renaissance period by Piero della Fransceca, Luca Pacioli, Leonardo da Vinci, Albrecht Duerer, 
Daniele Barbaro, and Johannes Kepler \cite{field1997}. Archimedean lattices are instantly appealing 
and they are seen in paintings and architecture. Indeed, these uniform Archimedean lattices 
are all around us: from patterns of household ceramic tiles, the weave in baskets (``kagom\'e'' 
means ``weave pattern'' in Japanese), and on to the atomic structures of materials. 

Quantum magnetic materials often demonstrate such regular patterns in the crystallographic 
structure of their magnetic atoms, which may interact via nearest-neighbour (NN) Heisenberg 
antiferromagnetic exchange interactions 
\cite{richter2004a,unjong2015,zheng2014}. Even the more exotic 
Archimedean lattices have been realized, see, 
e.g., CaV$_4$O$_9$ (CaVO) \cite{taniguchi1995}, SrCu$_2$(BO$_3$)$_2$ (SrCuBO) 
\cite{kageyama1999}, a polymeric iron(III) acetate (star) \cite{zheng2007} or 
Mx[Fe(O$_2$CCH$_2$)2NCH$_2$PO$_3$]$_6$nH$_2$O and 
Cu$_6$Al(SO$_4$)(OH)$_12$Cl3H$_2$O (maple-leaf) \cite{fennell2011}.  
Many of the Heisenberg antiferromagnets (HAFMs) on the Archimedean lattices 
are strongly ``frustrated,'' where frustration occurs when bonds compete with each other. 
For the systems studied here, this is due to an intrinsic incompatibility between the exchange 
interaction and the underlying lattice geometry. Most of the Archimedean lattices contain
triangles and are hence geometrically frustrated in the sense that not all pairs of NN spins
can be simultaneously antiparallel, as is otherwise favored by the HAFM exchange interaction
between pairs of spins.  
Indeed, the only unfrustrated cases here are for the square, honeycomb, SHD and CaVO lattices. 
Strong levels of such frustration can lead to novel states of quantum order or to states of magnetic 
disorder \cite{balents2010}. Such new physics that is driven by quantum mechanics is of immense interest to 
both theoreticians and experimentalists. Finally, the HAFM on the kagome lattice is 
of special importance  in the field of quantum magnetism because it provides an example of 
novel topological state of matter \cite{liao2017,yan2011,depenbrock2012}, which might be realized 
by the material herbertsmithite \cite{mendels2007}. The star lattice has much in common 
with the kagome lattice, including an infinitely degenerate ground state classically, although 
it has not been studied extensively. 
Although all sites on the Archimedean lattices are equivalent, not all NN bonds on these 
lattices necessarily have to be equivalent for all lattices. For example, bonds on the 
CaVO lattice in Fig. \ref{fig1} that lie on the squares are inequivalent to those bonds that connect 
these squares. This property gives us another criterion for dividing the Archimedean lattices 
into those lattices into which  NN bonds are all equivalent (i.e., square, honeycomb, triangle, 
and kagome) and the rest of the lattices where not all NN bonds are equivalent. However, 
frustration is undoubtedly the stronger influence on the behavior of these systems.

Previous studies \cite{ccm_high_order_archimedean_2014,richter2004a} considered the 
properties of the Heisenberg antiferromagnet on the Archimedean lattices by using a method of
quantum many-body theory called the coupled cluster method (CCM), although this 
was for the spin-half systems only. These analyses 
predicted (broadly) that three types of behavior occurs for the spin-half models: 
magnetically disordered systems (kagome and star); weakly ordered or possibly even 
disordered (maple-leaf, bounce, and trellis); and, magnetically ordered systems (square, 
honeycomb, CaVO, SHD, SrCuBO, and triangle).  Here we obtain accurate results for the 
ground-state energy and the order parameter for all Archimedean lattices and for spin quantum 
numbers, $s \le 4$. We provide results for scaling relations for 
both the ground-state energy and the order parameter as a function of the spin quantum 
number $s$, for each of the eleven lattices.

\section{Method}
The Hamiltonian of the HAFM is given by
\begin{equation}
H = J \sum_{\langle i,j \rangle } {\mathbf s}_i \cdot {\mathbf s}_j ~~,
\label{eq5}
\end{equation}
where ${\mathbf s}$ is the spin operator on the lattice site $i$, 
${\bf s}^2 = s ( s + 1 )$, and $J>0$. The symbol $\langle i,j \rangle$ indicates those bonds connecting 
adjacent sites (counting each bond once only). We consider this model here on all of the Archimedean lattices. The energy scale
is set by putting $J=1$ and the set of spin quantum numbers that we investigate here is $s\in \{\frac{1}{2}, 1,\frac{3}{2},\cdots,4\}$.  

The coupled cluster method (CCM)  \cite{refc1,refc2,refc3,refc4,refc5,refc6,refc7,refc8,refc9,ccmtheory1} 
is one of the most powerful and most versatile modern techniques available 
to us in quantum many-body theory. It has previously been applied very successfully to
various quantum magnetic systems. Details relating to the practical application of the CCM 
to these systems have been discussed extensively elsewhere (see e.g.,
 \cite{ccmtheory1, ccmtheory2,lattice_quantum_spin_systems_2010, ccm_high_order_square_triangle_1998,ccm_high_order_general_s_2001,ccm_high_order_square_2000,ccm_high_order_j1j2_square_2008}).
We remark that the CCM performs well also for those systems that 
are strongly frustrated.  The main alternative approximate methods either cannot 
be applied or are sometimes of limited usefulness only in such cases. For example, QMC 
techniques may demonstrate the sign problem for such systems and the ED method is limited in 
practice by computational power, especially for $s>1/2$, such that only very small 
lattices may be considered. Calculations for the density matrix renormalization (DMRG) 
method (and tensor-product methods) have generally been applied systems with lower 
spin quantum number as yet only in two spatial dimensions, although this method may 
also be applied in presence of strong frustration, in principle. 

We present a brief overview of CCM formalism before going on to
describe the some of the details only of the computational implementation of the 
CCM when it is applied to high-orders of approximation. Such computational 
methods are crucial to the accurate simulation of two-dimensional quantum 
magnetic systems. The highest level of approximation possible is limited only 
by the amount of computational resources available, as is described in an appendix for all
of the Archimedean lattices and for spin quantum number $s \le 4$. 
We begin any CCM calculation by choosing an appropriate 
{\it model} (or ``{\it vacuum}'')  state that is denoted by
$|\Phi \rangle$. The model state is the starting point for a CCM
calculation and we build in any additional quantum effects with 
respect to this state. A complete set of basis states may be obtained
by applying {\it creation} operators, $\{(C^+_L)\}$, to the 
model state. These creation operators are all mutually commuting. A 
detailed description of the possible choices of model state and 
associated choices of creation operators is given below for all
of the Archimedean lattices. The corresponding Hermitian 
adjoints of $\{(C^+_L)\}$ are denoted by $\{(C_L)\}$. These sets of
operators have the properties
\begin{equation}
\langle \Phi|C_L^+  = 0  = C_L|\Phi \rangle \quad \forall L\neq 0, \quad C_0^+\equiv 1 ~~ , 
\end{equation}
and
\begin{equation}
[C_L^+,C_J^+]_{-}=0=[C_L,C_J]_{-} ~~ .
\end{equation}
These operators are products of single-spin operators (see below) and so the index $L$ 
indicates a set of single-spin indices simultaneously. For the set $\{|\Phi\rangle,C_L^+\}$ one 
defines that
\begin{equation}
\langle\Phi|C_L C_J^+|\Phi\rangle=\delta_{LJ} ~~, 
\end{equation}
and
\begin{equation}
\sum_L C_L^+|\Phi\rangle\langle\Phi|C_L=1=|\Phi\rangle\langle\Phi|+\sum_{L\neq 0}C_L^+|\Phi\rangle\langle\Phi|C_L 
\end{equation}
which correspond to orthonormality and completeness, respectively. 
Again, the operators $C_L^+$ act upon the model state $|\Phi\rangle$, which can be understood as a generalized
vacuum state, and this approach allows for the construction of all possible states of the spin system. The CCM 
model states are often (though not always) given by the classical ground states of the lattice-spin systems 
under consideration. Thus, given a suitable model state and set of operators, ${\{|\Phi\rangle,C^+_L}\}$, the 
CCM parametrization of the exact ket and bra ground-state eigenvectors are given by
\begin{equation}
|\Psi\rangle=e^S|\Phi\rangle, \quad S=\sum_{L\neq 0}a_LC_L^+  \label{eq2.5}\\
\end{equation}
and
\begin{equation}
\langle \tilde{ \Psi}|=\langle \Phi |\tilde{S}e^{-S},  \quad \tilde{S}=1+\sum_{L\neq 0}\tilde{a}_LC_L~~,  
\end{equation}
for the ket and bra states, respectively. 
The operators $S$ and $\tilde{S}$ are correlation operators for the ket and bra state, 
where $\{a_L\}$ and $\{\tilde{a}_L\}$ are the corresponding correlation coefficients. 
The summation includes all possible configurations of the system. Note that the following 
normalizations are fulfilled also:
\begin{equation}
\langle\Phi|\Psi\rangle=\langle\tilde\Psi|\Psi\rangle=\langle\Phi|\Phi\rangle =1~~.
\end{equation}
The ground-state energy equation is found by multiplying the Schr\"odinger equation 
$H|\Psi\rangle = E|\Psi\rangle$ ($H$ is the Hamiltonian of the system) from the left with 
$\langle\Phi|e^{-S}$, and by using Eq. (\ref{eq2.5}), to give 
\begin{equation}\label{eq2.8}
e\equiv\frac{E}{N}=\frac{1}{N}\langle\Phi|e^ {-S} He^S |\Phi\rangle  ~~.
\end{equation}
The ground-state energy per spin $e$ depends on the ket-state correlation coefficients $\{a_L\}$ 
only. Furthermore, the similarity transformation in Eq. (\ref{eq2.8}) can be written as
the usual nested commutator expansion,
\begin{equation}\label{eq2.9}
e^{-S}He^S =H+ [H,S]_- +\frac{1}{2}[[H,S]_-,S]_- + \cdots  ~~ .
\end{equation}
Note also that this expression terminates after a finite set of terms if $H$ contains only a finite number 
of single-spin operators, as here. The ket- and bra-state coefficients $\{a_L\}$ and 
$\{\tilde{a}_L\}$ can be determined by requiring that $\bar{H}$ = $\langle\tilde{\Psi}|H|\Psi\rangle$ 
is stationary with respect to $\{\tilde{a}_L\}$ and $\{a_L\}$, respectively, which leads to the following equations
\begin{equation}\label{eq2.10}
\frac{\partial\bar{H}}{\partial\tilde{a}_L} = 0 \Leftrightarrow \langle\Phi|C_L e^{-S}He^S|\Phi\rangle =0, \forall L\not= 0, 
\end{equation}
and 
\begin{equation}\label{eq2.11}\frac{\partial\bar{H}}{\partial a_L} = 0 \Leftrightarrow 
\langle\Phi|\tilde{S} e^{-S}[H,C_L^+]_-e^S|\Phi\rangle =0, \forall L\not= 0~~.
\end{equation}
The ground state is specified completely once the coefficients {$\{a_L\}$ and $\{\tilde{a}_L\}$ 
have been found. Note  that Eq. (\ref{eq2.10}) is a nonlinear equation system in terms of 
the coefficients $\{a_L\}$ (ket equation system) and that Eq. (\ref{eq2.11}) is a linear equation system 
in terms of the coefficients $\{\tilde{a}_L\}$ (bra equation system). For every observable, it follows therefore that
\begin{equation}\label{eq2.12} \bar{A}=\langle\tilde{\Psi}|A|\Psi\rangle ={\bar A}(\{a_L,{\tilde a_L}\})~~.
\end{equation}
Here we use the order parameter $M/s$, which is defined by
\begin{equation}
\frac M s = -\frac 1{sN} \sum_{i=1}^N \langle \tilde \Psi | {\hat s}_i^z | \Psi \rangle,
\label{eq4}
\end{equation}
where we note that ${\hat s}_i^z$ is defined with respect to the local spin axes at site $i$ (after rotation of
the local spin axes) so that (notationally only) the spins in the model state appear to align 
in the negative $z$-direction (see also further below) at each site. If we set (trivially) all coefficients in $S$ and $\tilde S$ 
to zero then the ground-state wave function $|\Psi\rangle$ is given by the model state $|\Phi\rangle$, and 
so the order parameter, $M/s$, is equal to 1 for all lattices in this classical limit. The effect of quantum fluctuations 
is to reduce to the amount of magnetic order and so we expect $M/s < 1$ for all lattices. (We expect that $M/s = 1$ 
is true only in the ``classical'' limit, i.e., $s \rightarrow \infty$.) 

A discussion of the choice of model state for each lattice is given below, although we note again 
that a transformation of the local spin axes is used in all cases such that all spins point in the negative 
$z$-direction after ``rotation.'' This process allows us to treat all spins equivalently and it simplifies the 
mathematical solution of the CCM problem considerably. 
The corresponding operators are used therefore with respect to the CCM model state, such that 
\begin{equation}\label{eqn_rqz5}
 |\Phi\rangle=\bigotimes_{i=1}^N|\downarrow\rangle_i, \hspace{0.2cm} C_I^+  =s_{i_1}^+, s_{i_1}^+s_{i_2}^+, s_{i_1}^+s_{i_2}^+s_{i_3}^+, \cdots ~~ .
\end{equation}
where $i_1$,$i_2$ and $i_3$ denote arbitrary lattice sites.
The CCM formalism would be exact if all possible multi-spin cluster
correlations could be included in $S$ and $\tilde S$, although  this is normally 
impossible to achieve practically. In most cases, the following approximation schemes 
within $S$ and $\tilde{S}$ are used, namely:
\begin{itemize}
\item The SUB$n$ scheme: all correlations involving only $n$ or
fewer spins are retained, although no further restriction is made concerning their spatial
separation on the lattice; 
\item The SUB$n$-$m$ scheme:  all SUB$n$ correlations
spanning a range of no more than $m$ adjacent lattice sites are retained; and
\item The localized LSUB$n$ scheme: all multi-spin correlations over
all distinct locales on the lattice defined by $n$ or fewer contiguous sites are retained.
\end{itemize}
Note that the LSUB$n$ and SUB$m$-$m$ schemes are identical for the limiting case when 
$s = 1/2$ and $n = m$. For higher spins $s$, the LSUB$n$ scheme is equivalent to the 
SUB$n$-$m$ scheme if and only if $n = 2 \cdot s \cdot m$. Note that we always use the 
SUB$m$-$m$ scheme in this article. 
We see that these approximation schemes allow us to increase the level of approximation in a
systemic and well-controlled manner. Furthermore, we can also attempt to extrapolate 
our ``raw" SUB$n$, SUB$n$-$m$, and LSUB$n$ results in the limits $n,m \rightarrow
\infty$ in order to obtain even more accurate results. Although there is no general theory 
for the scaling behavior $m\rightarrow\infty$, there is nevertheless much empirical evidence 
relating to how to extrapolate the raw SUB$m$-$m$ data 
\cite{ccm_high_order_square_triangle_1998,ccm_high_order_square_2000,ccm_high_order_j1j2_square_2008,ccm_high_order_general_s_2001}. 
For example, the following scheme is generally used for the ground-state energy per spin $e$,
\begin{equation}
 \label{eqn_extra_1}
 e(m)=a_0+a_1 m^{-2}+a_2 m^{-4} ~~.
\end{equation}
By contrast, two commonly used extrapolation schemes for the order parameter $M$ are given by
\begin{equation}
 M_I(m)=b_0+b_1 m^{-1}+ b_2 m^{-2}  \label{eqn_extra_2}
\end{equation}
and 
\begin{equation}
 M_{II}(m)=c_0+c_1 m^{-0.5}+ c_2 m^{-1.5} ~~, \label{eqn_extra_3}
\end{equation}
referred to as schemes I and II, respectively. 
Extrapolation scheme II is arguably more favorable for those systems that 
exhibit a magnetically disordered ground state and / or an
order to disorder transition. 
The quality of the extrapolation is generally improved by omitting the lowest level 
of approximation, typically LSUB2 or SUB2-2, which is carried out only if enough data points for extrapolation 
are available. Furthermore, note that we extrapolate results for even values of $m$ for the lattices: square, SHD, 
honeycomb, SrCuBO, and CaVO. By contrast, we use both results from both odd and even values of $m$ for all 
other lattices because the ``odd/even'' staggering effect \cite{farnell2008}
is much less pronounced for these frustrated systems, especially 
at higher orders of SUB$m$-$m$ approximation.

The model states used here are the classical ground states (i.e., products of independent single-spin
states) for all lattices. For example, the model state for all of the bipartite 
(unfrustrated) Archimedean lattices is the collinear (N\'eel) classical ground state in 
which nearest neighbor spins are antiparallel. The bipartite lattices are given by:
square ($z=4$) \cite{ccm_high_order_square_triangle_1998,ccm_high_order_square_2000,ccm_high_order_j1j2_square_2008},  
honeycomb ($z=3$) \cite{ccm_high_order_honeycomb_2011}, 
CaVO ($z=3$) \cite{ccm_high_order_CaVO_2005},  and
SHD ($z=3$) \cite{ccm_high_order_archimedean_2014}. 
($z$ is the coordination number of the lattice.) 
The frustrated SrCuBO lattice ($z=5$) uses a model state in which 
nearest-neighbor spins are antiparallel on the squares only for this 
lattice shown in Fig. \ref{fig1}. 
\cite{ccm_high_order_SrCuBO_2005}.  An explicit restriction is imposed 
on the creation operators $\{C_I^+\}$ in $S$ for these model states, namely, 
that relationship $s^z_T= \sum_i s^z_i=0$ in the original (unrotated) 
spin coordinates is preserved. This restriction 
guarantees that the approximate CCM ground-state wave function lies in the 
correct ($s^z_T=0$) subspace for these lattices. The classical ground-state 
energy per bond for the HAFM on the square, honeycomb, CaVO, and SHD lattices 
is given by $-s^2$, whereas the classical ground-state energy per bond on the SrCuBO 
lattice is given by $-0.6s^2$.  
The model state for the triangular lattice ($z=6$) \cite{ccm_high_order_square_triangle_1998} 
is given by spins on nearest-neighboring sites that form angles 
of 120$^\circ$ to each other. There is no restriction on $s^z_T$ in this case and 
the classical ground-state energy per bond is given by $-0.5s^2$. 
The model states for the maple-leaf ($z=5$) \cite{ccm_high_order_maple_bounce_2011}, 
trellis ($z=5$) \cite{ccm_high_order_archimedean_2014}, 
and bounce ($z=4$) \cite{ccm_high_order_maple_bounce_2011} 
lattices are the classical ground states, and again there is no restriction on 
$s^z_T$. For both the maple-leaf and bounce lattices, the classical
ground state comprises a six-sublattice struture with a specified pitch
angle. By contrast, for the trellis lattice the classical ground state
comprises incommensurate spirals along a chain direction. In all three
cases, we take into account that quantum fluctuations can lead to
a characteristic spiral or pitch angle that is different from its
classical counterpart. Hence, in these three cases the
characteristic angle is taken as a free parameter, which is selected in 
practice to minimize the respective ground-state energy obtained 
separately at level of SUB$m$-$m$ level of approximation. 
The classical ground-state energy per bond is now given by 
$-s^2(1+\sqrt{3})/5$ (maple-leaf) \cite{ccm_high_order_maple_bounce_2011}, 
$-0.65s^2$ (trellis) \cite{richter2004a}, 
and $-0.75s^2$ (bounce) \cite{ccm_high_order_maple_bounce_2011}.
Finally, the case for the kagome ($z=4$) \cite{ccm_high_order_triangle_kagome_2001,ccm_high_order_kagome_square_2010,goetze2011}
and star ($z=3$) \cite{richter2004a,ccm_high_order_archimedean_2014,richter2004b} lattices is slightly more complicated 
because there are an infinite number of possible classical ground states to choose 
from potentially, where each potential classical ground state has
nearest-neighboring spins on the ``triangles'' of these lattices that 
form angles of 120$^{\circ}$ to each other. 
(The relative angles of those spins connecting triangles on the star lattice
is 180$^{\circ}$.) 
Here  we consider the ``$\sqrt{3}\times\sqrt{3}$'' and ``$q=0$'' states only. 
No restriction on $s^z_T$ is imposed for either the kagome or star lattices 
and the classical ground-state energy per bond is given by $-0.5s^2$
and $-2s^2/3$ for the HAFM on the kagome and star lattices,
respectively. The number of fundamental clusters that 
are distinct under the translational and point-group symmetries of the lattice
and Hamiltonian (after rotation of local spin axes) and also that preserves 
$s_T^z$ (as appropriate) is denoted $N_f=N_f(m)$ at a given level of 
SUB$m$-$m$ approximation. The size of the CCM equations (measured 
in terms of the number of terms and also memory usage in bytes) at a given level of 
SUB$m$-$m$ approximation is denoted $N_t=N_t(m)$.

CCM equations may be derived and solved analytically at low orders of approximation. A full 
explanation of how this is carried out for the SUB2-2 approximation for the spin-half {\it XXZ}
model on the square lattice, for example, is given on pages 117 to 122 of Ref. \cite{lattice_quantum_spin_systems_2010}. 
However, highly intensive computational methods
\cite{ccm_high_order_square_triangle_1998,ccm_high_order_square_2000,ccm_high_order_j1j2_square_2008,ccm_high_order_general_s_2001}
rapidly become essential at higher orders of SUB$m$-$m$ approximation because the number 
of fundamental clusters (and so therefore also the computational resources necessary to store and 
solve them) scales approximately exponentially with $m$. There are four distinct steps to carrying out 
high-order CCM calculations for the ground state and an efficient computer code has been developed \cite{code}. 
The first step is to enumerate all ($N_f$) fundamental clusters that are distinct under the lattice and Hamiltonian 
symmetries (and perhaps that also satisfy $s_T^z=0$) at a specific level of SUB$m$-$m$ approximation and 
for a given lattice and spin quantum number $s$. 
The second step involves finding and storing the basic CCM ground- and excited-state equations. 
The third step involves solving both the ground- and excited-state equations and to obtain the 
macroscopic quantities for these states. A minimal ground-state energy solution as fnction 
of some explicit ``angle'' might be necessary if a ``spiral'' or ``canted'' model state \cite{zinke2009} 
is used, as mentioned above. 
The fourth step is to create and input a 
basic  script that defines completely the problem to be solved by the CCM code. 
Separate scripts were written for all eleven lattices considered here. The computational
resources necessary to carry out high-order CCM calculations are considered in an 
Appendix.

\section{Results}


Extrapolated results for the ground-state energy $E_g/(NJs^2)$ of the spin-half 
Heisenberg model on all of the lattices and for all values of spin quantum number, 
$s$, are given in Table \ref{tab1}. Inspection of the extrapolated ground-state 
energy indicates that the $\sqrt{3}\times\sqrt{3}$ state is favored over the 
$q = 0$ state to form the ground state (i.e., it has lower ground-state energy) for the kagome and 
star lattices for most values of the spin quantum number, $s$. However, we 
find that the $q = 0$ state is favored over the $\sqrt{3}\times\sqrt{3}$ state 
for the kagome lattice for the specific case of $s = 1/2$ and it is favored over 
the $\sqrt{3}\times\sqrt{3}$ state for the star lattice for $s = 1/2$ and $s = 1$. 
As seen previously \cite{ccm_high_order_archimedean_2014}, CCM results for
the ground-state energy per bond (in units of $J$) for the spin-half HAFM (i.e., 
$E_g/(JN_b)$: square, $-$0.3348; honeycomb, $-$0.3631; CaVO, $-$0.3689; SHD, 
$-$0.3702; SrCuBO, $-$0.2312; triangle, $-$0.1843; bounce, $-$0.2824; trellis, 
$-$0.2416; maple-leaf, $-$0.2124; star, $-$0.3110; kagome, $-$0.2179) compare well to 
those results of other approximate methods (i.e., $E_g/(JN_b)$: square, $-$0.3347 
\cite{sandvik1997}; honeycomb, $-$0.3630 \cite{reger1989}; CaVO, $-$0.3691 
\cite{jagannathan2006}; SHD, $-$0.3688 \cite{tomczak2001}; SrCuBO, $-$0.23 
to $-$0.24 \cite{corboz2013}; triangle, $-$0.1823 \cite{chernyshev2009}; 
bounce, $-$0.2837 \cite{richter2004a}; trellis, $-$0.2471 \cite{richter2004a,richter2004b}; 
maple-leaf, $-$0.2171 \cite{richter2004a}; star, $-$0.316 to $-$0.318 
\cite{yang2010}; kagome, $-$0.21876 to $-$0.2193 
\cite{liao2017,yan2011,depenbrock2012,lauchli2011}). ($N_b$ is the number 
of lattice bonds.) 

Isolated results exist only for certain Archimedean lattices when we 
set $s>1/2$. CCM results for the ground-state energy of the spin-one, 
square-lattice HAFM of $E_g/(NJs^2)=-$2.32856 compare well to results of second-order
spin-wave theory \cite{weihong1991a} of $-$2.3284, third-order spin-wave theory 
\cite{hamer1992} of $-$2.32815,  and series expansions \cite{weihong1991a} of 
$-$2.3279(3). CCM results for the ground-state 
energy of the spin-one, honeycomb-lattice HAFM of $E_g/(NJs^2)=-$1.83061
are in good agreement with results of spin-wave theory \cite{weihong1991b,oitmaa1991} of 
$-$1.8313 and results of series expansions \cite{oitmaa1991} of 
$-$1.8278(8). CCM results for the ground-state energy for the spin-one, 
kagome-lattice HAFM  of $E_g/(NJs^2)=-$1.40315 correspond well to
results of Ref. \cite{changlani2015} of $-$1.410(2), 
Ref. \cite{liu2015} of $-$1.4109(2), and Ref. \cite{nishimoto2015}
of $-$1.41095. CCM results for the ground-state energy for the 
$s=3/2$, kagome-lattice HAFM  of $E_g/(NJs^2)=-$1.26798 compare well 
to results of Ref. \cite{liu2016} of $-$1.265(2). (Note that results for 
the scaling of the ground-state energy with spin quantum number $s$ are also 
compared to results of spin-wave theory below, where they exist for the 
Archimedean lattices.)

Extrapolated results for the order parameter, $M/s$, for the HAFM on 
all of the lattices and for all values of spin quantum number, $s$, are given 
in Table \ref{tab2}. We see that the unfrustrated models (i.e.: square, 
honeycomb, SHD, and CaVO) and two frustrated lattices (i.e.: triangle 
and SrCuBO) are magnetically ordered for all values of the spin quantum 
number, $s$, including the limiting case of the spin-half systems. The HAFM 
on the trellis, bounce, and maple-leaf lattices is either disordered or very weakly 
magnetically ordered \cite{ccm_high_order_archimedean_2014} for the spin-half 
system, and so results of both extrapolation schemes I and II of Eqs. (\ref{eqn_extra_2}) and 
(\ref{eqn_extra_3}) are given in Table \ref{tab2} for $s = 1/2$ and for these lattices. By 
contrast, the HAFM is magnetically ordered for $s > 1/2$ for these lattices and so  
the extrapolation scheme I of Eq. (\ref{eqn_extra_2}) is used for $s > 1/2$. 
Results of both extrapolation schemes I and II of Eqs. (\ref{eqn_extra_2}) and (\ref{eqn_extra_3}) 
are given in Table \ref{tab2}  for the kagome and star lattices. Differences between 
the two extrapolation schemes I and II are seen in Table  \ref{tab2} for both the star 
lattice and (as noted previously in Ref. \cite{goetze2011}) also for the kagome lattice.  
However, it is clear that the kagome and star lattices are predicted to be disordered 
(i.e., $M/s=0$) by using both extrapolation schemes of Eqs. (\ref{eqn_extra_2}) and 
(\ref{eqn_extra_3}) for $s=1/2$ \cite{ccm_high_order_archimedean_2014}. 
By contrast, the kagome and star lattices are magnetically disordered 
(or weakly magnetically ordered) for $s=1$. The kagome lattice is either weakly ordered 
or magnetically ordered for $s=3/2$. Furthermore, it is clear also that all other results for 
higher $s$ for the star ($s>1$) and kagome ($s>3/2$) lattices are magnetically ordered 
from both extrapolation schemes I and II. 

As seen previously  \cite{ccm_high_order_archimedean_2014}, CCM results 
for the order parameter $M/s$ for the spin-half HAFM (i.e.: square, 0.619; honeycomb, 
0.547; CaVO, 0.431; SHD, 0.366; SrCuBO, 0.404; triangle, 0.373; bounce, 0 to 
0.122; trellis, 0.0 to 0.040; maple-leaf, 0 to 0.178; star, 0; kagome, 0) are 
again found to compare well to results of other approximate methods (i.e.: 
square, 0.614 \cite{sandvik1997}; honeycomb, 0.535 \cite{castro2006}; 
CaVO, 0.356 \cite{troyer1996}; SHD, 0.509 \cite{tomczak2001}; SrCuBO, 
0.42 \cite{corboz2013}; triangle, 0.410 \cite{white2007}; bounce, 0.268 
\cite{richter2004a}; trellis, 0.222 \cite{richter2004a}; maple-leaf \cite{richter2004a},
0.218; star, 0.094 to 0.15 \cite{yang2010}; kagome, 0 
\cite{liao2017,yan2011,depenbrock2012,lauchli2011}).  

Again, isolated results exist only for the order parameter for certain 
Archimedean lattices with $s>1/2$. CCM results for the order parameter 
of the spin-one, square-lattice HAFM of $M/s=$0.79942 compare well to 
spin-wave theory \cite{weihong1991a} of 0.8034, third-order spin-wave theory
of \cite{hamer1992} of 0.80427, and series expansions \cite{weihong1991a} of 0.8039(4), as  
well as recent results from infinite projected entangled pair states (iPEPS) \cite{niesen2017} 
of $M/s=$0.802(7). CCM results for the order parameter of the spin-one, 
honeycomb-lattice HAFM of $M/s=$0.74123 are in good agreement with results of 
spin-wave theory of \cite{weihong1991b,oitmaa1991} of $M/s=$0.7418 and  
results of series expansions \cite{oitmaa1991} of 0.748(3). Results of other 
approximate methods for the spin-one, kagome-lattice HAFM suggest that there is 
no magnetic long-range order \cite{changlani2015,liu2015,nishimoto2015,oitmaa2016}, 
although Ref. \cite{cepas2011} indicated $\sqrt{3}\times\sqrt{3}$
ground-state long-range order for integer spin quantum numbers, including 
$s = 1$. Previous CCM calculations \cite{goetze2015} for general spin quantum number 
concluded that CCM results were consistent with magnetic disorder for the spin-one, 
kagome-lattice HAFM. Series expansion calculations \cite{oitmaa2016} indicated that $M/s=0.14 \pm 0.03$ 
for the $s=3/2$ kagome-lattice HAFM and recent tensor network calculations \cite{liu2016} 
also indicate that the $s=3/2$ system is $\sqrt{3}\times\sqrt{3}$ long-range ordered.
Indeed, the consensus from approximate methods 
\cite{goetze2011,cepas2011,goetze2015,chernyshev2014} is that the kagome-lattice HAFM 
demonstrates $\sqrt{3}\times\sqrt{3}$ ground-state long-range order for $s\ge3/2$. 
(Note that results for the scaling of the order parameter with spin quantum number 
$s$ are also compared to results of spin-wave theory below, where they exist for the 
Archimedean lattices.)

\begin{table}
\caption{Extrapolated SUB$m$-$m$ results for the ground-state energy $E_g/(NJs^2)$ for the HAFM on all Archimedean lattices and with $s \in \{1/2, 1, 3/2, \cdots ,4 \}$. (The extrapolation scheme of Eq. (\ref{eqn_extra_1}) is used with (a) only even approximations with SUB4-4 or higher or $(b)$ all data with SUB4-4 or higher. Results for the kagome and star lattices are shown for the $\sqrt{3}\times\sqrt{3}$ model state, except for $s=1/2$ for the kagome lattice and $s=1/2,1$ for the star lattice where the results for the $q=0$ state are shown.)}
\begin{center}
\begin{tabular}{|l|c|c|c|c|c|c|c|c|}  \hline
            &$s=1/2$ &$s=1$  &$s=3/2$ &$s=2$ &$s=5/2$ &$s=3$ &$s=7/2$ &$s=4$         \\ \hline
Square$^{(a)}$ 		&$-$2.67857	&$-$2.32856	&$-$2.21633	&$-$2.16119	&$-$2.12844	&$-$2.10675	&$-$2.09135	&$-$2.07981  \\ \hline
Honeycomb$^{(a)}$  &$-$2.17885	&$-$1.83061	&$-$1.71721	&$-$1.66159	&$-$1.62862	&$-$1.60681	&$-$1.59133	&$-$1.57976  \\ \hline
CaVO$^{(a)}$  		&$-$2.21307	&$-$1.84297	&$-$1.72497	&$-$1.66724	&$-$1.63306	&$-$1.61048	&$-$1.59444	&$-$1.58249  \\ \hline
SHD$^{(a)}$  		&$-$2.22133	&$-$1.84592	&$-$1.72697	&$-$1.66871	&$-$1.63421	&$-$1.61142	&$-$1.59524	&$-$1.58317  \\ \hline
SrCuBO$^{(a)}$  	&$-$2.31212	&$-$1.90314	&$-$1.76867	&$-$1.70159	&$-$1.66135	&$-$1.63452	&$-$1.61535	&$-$1.60097  \\ \hline
Triangle$^{(b)}$  	&$-$2.21117	&$-$1.84097	&$-$1.72421	&$-$1.66697	&$-$1.63303	&$-$1.61057	&$-$1.59461	&$-$1.58268  \\ \hline
Bounce$^{(b)}$  	&$-$2.25904	&$-$1.85797	&$-$1.73423	&$-$1.67405	&$-$1.63845	&$-$1.61493	&$-$1.59823	&$-$1.58578  \\ \hline
Maple-Leaf$^{(b)}$  &$-$2.12376	&$-$1.72481	&$-$1.60159	&$-$1.54135	&$-$1.50619	&$-$1.48246	&$-$1.46559	&$-$1.45300  \\ \hline
Trellis$^{(b)}$  		&$-$2.41580	&$-$1.98823	&$-$1.86186	&$-$1.80073	&$-$1.76467	&$-$1.74142	&$-$1.72447	&$-$1.71183  \\ \hline
Kagome$^{(b)}$  	&$-$1.74345	&$-$1.40315	&$-$1.26798	&$-$1.20260	&$-$1.16272	&$-$1.13598	&$-$1.11680	&$-$1.10237 \\ \hline
Star$^{(b)}$   		&$-$1.86622	&$-$1.39597 &$-$1.26002	&$-$1.19494	&$-$1.15595	&$-$1.13002	&$-$1.11152	&$-$1.09766  \\ \hline
\end{tabular}
\end{center}
\vspace{20pt}
\label{tab1}
\end{table}

\begin{table}
\caption{Extrapolated SUB$m$-$m$ results for the order parameter $M/s$ for the Heisenberg antiferromagnet  on all Archimedean lattices and with $s \in \{1/2, 1, 3/2, \cdots ,4 \}$. (The extrapolation scheme I of Eq. (\ref{eqn_extra_2}) is used with $(c)$ only even approximations with SUB4-4 or higher or (d) all data with SUB4-4 or higher; (e) extrapolation scheme II of Eq. (\ref{eqn_extra_3}) is used with all data with SUB4-4 or higher. Results for the kagome and star lattices are shown for the $\sqrt{3}\times\sqrt{3}$ model state, except for $s=1/2$ for the kagome lattice and $s=1/2,1$ for the star lattice where the results for the $q=0$ state are shown.)}
\begin{center}
\begin{tabular}{|l|c|c|c|c|c|c|c|c|}  \hline
            &$s=1/2$ &$s=1$  &$s=3/2$ &$s=2$ &$s=5/2$ &$s=3$ &$s=7/2$ &$s=4$  \\ \hline
Square$^{(c)}$		&0.61862	&0.79942	&0.86634	&0.90026	&0.92053	&0.93396	&0.94340	&0.95056 \\ \hline
Honeycomb$^{(c)}$	&0.54730	&0.74123	&0.82492	&0.86894	&0.89552	&0.91319	&0.92576	&0.93514 \\ \hline
CaVO$^{(c)}$		&0.43058	&0.70738	&0.80302	&0.85292	&0.88297	&0.90292	&0.91708	&0.92744 \\ \hline
SHD$^{(c)}$			&0.36647	&0.68631	&0.78847	&0.84139	&0.87356	&0.89504	&0.91033	&0.92174 \\ \hline
SrCuBO$^{(c)}$		&0.40354	&0.64239	&0.72799	&0.77385	&0.80378	&0.82563	&0.84264	&0.85642 \\ \hline
Triangle$^{(d)}$		&0.37251	&0.68850	&0.79845	&0.85318	&0.88487	&0.90528	&0.91946	&0.92988 \\ \hline
Bounce$^{(d)}$		&0.12208 (0$^{(e)}$)	&0.60091	&0.73566	&0.80221	&0.84257	&0.86956	&0.88877	&0.90310 \\ \hline
Maple-leaf$^{(d)}$	&0.17816 (0$^{(e)}$)	&0.60681	&0.73121	&0.79685	&0.82569	&0.85387	&0.87436	&0.88988 \\ \hline
Trellis$^{(d)}$		&0.03980 (0$^{(e)}$)	&0.61876	&0.75301	&0.81817	&0.85670	&0.87450	&0.89304	&0.90710 \\ \hline
Kagome$^{(d)}$		&0	&0.15933	&0.41742 &0.49975	&0.55729	&0.59860	&0.63018	&0.65536 \\ \hline
Kagome$^{(e)}$		&0    &0     	&0.07439	&0.20290	&0.29425	&0.35825	&0.40623	&0.44397 \\ \hline
Star$^{(d)}$			&0	&0.19158	&0.42553	&0.50031	&0.54444	&0.57197	&0.58997	&0.60233 \\ \hline
Star$^{(e)}$			&0    &0       	&0.22317	&0.35883	&0.45135	&0.51725	&0.56665	&0.60521 \\ \hline
\end{tabular}
\end{center}
\vspace{20pt}
\label{tab2}
\end{table}

The scaling laws of the ground-state energy and the order parameter are shown in 
Table \ref{tab3} and they are illustrated by Figs. \ref{fig2} and \ref{fig3}, respectively. 
By allowing the index $\nu$ to vary as an explicit parameter for the 
ground-state energy in  
\begin{equation}
\frac {E_g}{NJs^2} = \frac {E_{\rm cl}}{NJs^2} + \alpha s^{\nu} ~~,
\end{equation} 
we found that $E_g/(NJs^2)$ was found to scale with $s^{-1}$ to leading order for 
all lattices. The classical result $E_{\rm cl}/(NJs^2)$ is 
known to be correct in the asymptotic limit $s \rightarrow \infty$ and so this 
expression is used explicitly in the scaling relations. As in Ref. \cite{goetze2011}, 
the extreme (quantum) cases of $s=1/2$ and $s=1$ are not used in fitting the data in 
Table \ref{tab1} to the scaling relations (i.e., $s = \{3/2, 2, 5/2, 3, 7/2, 4\}$ are used here). 
The asymptotic relation for the ground-state energy to second order in 
$s$ is therefore given by 
\begin{equation}
\frac {E_g}{NJs^2} = \frac {E_{\rm cl}}{NJs^2} + \alpha s^{-1} + \beta s^{-2} ~~, 
\end{equation} 
and the coefficients $\alpha$ and $\beta$ are presented in this 
Table \ref{tab3} for all lattices. Results for $E_g/(NJs^2)$ plotted as a 
function of $s^{-1}$ are shown in Figs. \ref{fig2}. Associated line fits to the data 
using the values for $\alpha$ and $\beta$ in Table \ref{tab3} are shown also.

The index $\mu$ was allowed to vary as an 
explicit parameter for the order parameter in 
\begin{equation}
\frac Ms = 1+ \gamma s^{\mu} ~~.
\end{equation} 
Note that the classical result $M/s=1$ in the limit $s \rightarrow \infty$ is 
assumed explicitly in this equation (as in Ref. \cite{goetze2011}) and that the 
extreme (quantum) cases of $s=1/2$ and $s=1$ are not used in fitting the data 
to this scaling relation (i.e., $s = \{3/2, 2, 5/2, 3, 7/2, 4\}$ are used here). 
$M/s$ was found to scale with $s^{-1}$ to leading order for most lattices.  
However, it was found that $M/s$ scales with $s^{-1/2}$ to first order for 
the kagome lattice (as seen previously in Ref. \cite{goetze2011}),
$s^{-1/3}$ to first order for the star lattice, 
and $s^{-2/3}$ to first order for the SrCuBO lattice. 
The results for the exponent of the star lattice are the least reliable here, which also
shows the difficulty of simulating this system. For example, results for such line fits 
for the star lattice using data for $s = \{2,5/2,3,7/2,4\}$, $s = \{5/2,3,7/2,4\}$, and 
$s = \{3,7/2,4\}$, respectively, suggest that the exponent for the order parameter 
might even have a magnitude that is less than $1/3$. It is clear though that $M/s$ 
does not scale with $s^{-1}$ for the star lattice and that it is the 
most extreme case studied here. The asymptotic relation for the order parameter $M/s$ to 
second order with $s$ is therefore given by 
\begin{equation}
\frac Ms  = 1 +  \gamma s^{-1} + \delta s^{-2} ~~,
\end{equation} 
for most lattices. 
(Appropriate relations to second order were used for the 
kagome, star and SrCuBO lattices, as described in Table \ref{tab3}.)
Results for $\gamma$ and $\delta$ are shown in Table \ref{tab3}. 
Results for $M/s$ plotted as a function of $s^{-1}$ are shown in Fig. \ref{fig3}.
Associated line fits to the data 
using the values for $\gamma$ and $\delta$ in Table \ref{tab3} are shown also.
For the sake of clarity, only results for the order parameter 
obtained via extrapolation scheme I of Eq. (\ref{eqn_extra_2}) are included in 
Fig. \ref{fig3} for the kagome and star lattices. 

CCM results on the square lattice of $\alpha=-0.3161$ and $\gamma=-0.1961$  
for the ground-state energy and order parameter, respectively, compare well to first order 
with results of spin-wave theory (SWT) \cite{hamer1992} of  $\alpha=-0.315895$ and 
$\gamma = -0.1966019$. CCM results on the honeycomb lattice of $\alpha=-0.3151$ and 
$\gamma=-0.2583$  for the ground-state energy and order parameter, respectively, 
also compare generally well to first order with results of SWT \cite{weihong1991b} of 
$\alpha = -0.31476$ and $\gamma = -0.2582$, respectively. CCM results on the triangular 
lattice of $\alpha=-0.3272$ and $-0.2666$ for the ground-state energy and order parameter, 
respectively, yet again compare well to first order with results of SWT \cite{chernyshev2009} 
of $\alpha=-0.32762$ and $\gamma=-0.261303$ (see also Ref. \cite{chubukov1994}). 
Scaling results for the honeycomb lattice agree well with previous CCM results \cite{bishop2016}, 
although higher orders of approximation have been achieved here. 
Results for the kagome lattice from self-consistent spin-wave theory \cite{chubukov1992} state that 
the order parameter scales with $s^{-2/3}$. CCM results for $M/s$ for kagome lattice for 
the $q=0$ model state were found previously \cite{goetze2011} to scale with $s^{-2/3}$. 
CCM results for $M/s$ for the star lattice for the $q=0$ model state are also found 
to scale with $s^{-2/3}$. 
Note again that the $\sqrt{3}\times\sqrt{3}$ model state has lower energy than the $q=0$ state for 
$s \ge 1$ for the kagome lattice and $s \ge 3/2$ for the star lattice. 
In accordance with previous work \cite{goetze2011} we remember again that 
$M/s$ for the $\sqrt{3}\times\sqrt{3}$ ground state scales with $s^{-1/2}$ for the kagome 
lattice, rather than $s^{-2/3}$ as for the $q=0$ state.  It is this form of scaling that is shown in 
in Table \ref{tab3} and in Fig. \ref{fig3} for the kagome lattice. 
No scaling information with $s$ exists from other approximate methods for the 
CaVO, SHD, SrCuBO, maple-leaf, bounce, trellis, or star lattices, as far as we are aware. 


\begin{table}
\caption{Scaling of the ground-state energy in Table \ref{tab1} and the order parameter in 
Table \ref{tab2} with respect to $s$. (Scaling for the ground-state energy uses the relation: 
$E_g/(NJs^2) = E_{\rm cl}/(NJs^2) + \alpha s^{-1} + \beta s^{-2}$. Scaling for the order parameter use 
the relations:  scheme (1) uses $M/s = 1 +  \gamma s^{-1} + \delta s^{-2}$; scheme (2) uses 
$M/s = 1 + \gamma s^{-2/3} + \delta s^{-4/3}$; scheme (3) uses $M/s = 1 + \gamma 
s^{-1/2} + \delta s^{-1}$;  and scheme (4) uses $M/s = 1 + \gamma s^{-1/3} + \delta s^{-2/3}$. 
All fits of the data to these relations were carried out using $s \ge 1.5$ 
(i.e., $s = \{3/2, 2, 5/2, 3, 7/2, 4\}$ are used here. Results are shown for the two extrapolation 
schemes I $(d)$ and II $(e)$ in Table \ref{tab2} for the kagome and star lattices.)}
\begin{center}
\begin{tabular}{|l|c|c|c|c|c|}  \hline
			&\multicolumn{3}{c|}{$E_g/(NJs^2)$}	&\multicolumn{2}{c|}{$M/s$}  			\\ \hline 
			&$E_{\rm cl}/(NJs^2)$ &$\alpha$			&$\beta$		&$\gamma$		&$\delta$			\\ \hline 
Square		&$-2$			&$-$0.3161		&$-$0.0126	&$-$0.1961$^{(1)}$		&$-$0.0066$^{(1)}$	\\ \hline 
Honeycomb	&$-1.5$			&$-$0.3151		&$-$0.0162	&$-$0.2583$^{(1)}$ 		&$-$0.0067$^{(1)}$ 	\\ \hline 
CaVO		&$-1.5$			&$-$0.3254		&$-$0.0181	&$-$0.2873$^{(1)}$		&$-$0.0125$^{(1)}$ 	\\ \hline 
SHD			&$-1.5$			&$-$0.3281		&$-$0.0186	&$-$0.3123$^{(1)}$		&$-$0.0080$^{(1)}$	\\ \hline
SrCuBO		&$-1.5$			&$-$0.4042		&0.0018		&$-$0.3708$^{(2)}$		&0.0186$^{(2)}$	\\ \hline 
Triangle		&$-1.5$			&$-$0.3272		&$-$0.0137	&$-$0.2666$^{(1)}$		&$-$0.0536$^{(1)}$	\\ \hline 
Bounce		&$-1.5$			&$-$0.3382		&$-$0.0197	&$-$0.3857$^{(1)}$		&$-$0.0171$^{(1)}$	\\ \hline 
Maple-leaf	&$-1.36603$		&$-$0.3450		&$-$0.0124	&$-$0.4679$^{(1)}$		&0.1008$^{(1)}$		\\ \hline 
Trellis		&$-1.625$		&$-$0.3421		&$-$0.0195	&$-$0.3678$^{(1)}$		&$-$0.0006$^{(1)}$	\\ \hline  
Kagome		&$-1$			&$-$0.4140		&0.0180		&$-$0.6507$^{(3)}$$^{(d)}$	&$-$0.0779$^{(3)}$$^{(d)}$	\\ \hline 
			&				&				&			&$-$1.0676$^{(3)}$$^{(e)}$	&$-$0.0810$^{(3)}$$^{(e)}$	\\ \hline 
Star			&$-1$			&$-$0.3903		&0.0005		&$-$0.5293$^{(4)}$$^{(d)}$	&$-$0.1366$^{(4)}$$^{(d)}$	\\ \hline 	
			&				&				&			&$-$1.0576$^{(4)}$$^{(e)}$	&$-$0.0980$^{(4)}$$^{(e)}$	\\ \hline 
	
\end{tabular}
\end{center}
\vspace{20pt}
\label{tab3}
\end{table}



\begin{figure}
\begin{center}
\epsfxsize=13cm
\centerline{\epsffile{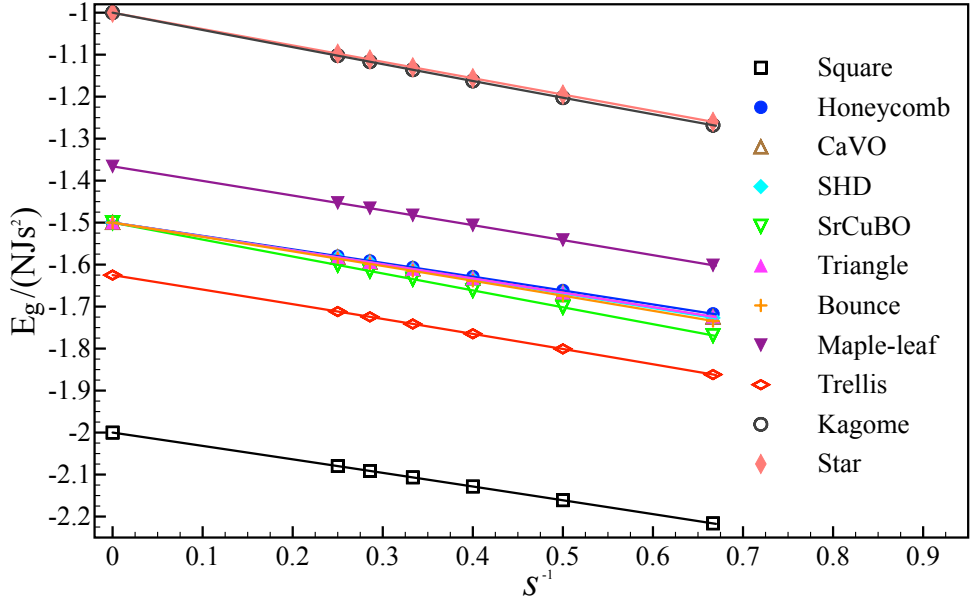}}
\caption{\label{fig2}(Color online) 
Results for the ground-state energy per site, $E_g/(NJs^2)$ plotted as a function of $s^{-1}$ for $s \ge 3/2$. 
(Associated line fits to the data using the relations outlined in Table \ref{tab3} are shown also.)}
\end{center}
\end{figure}

\begin{figure}
\begin{center}
\epsfxsize=13cm
\centerline{\epsffile{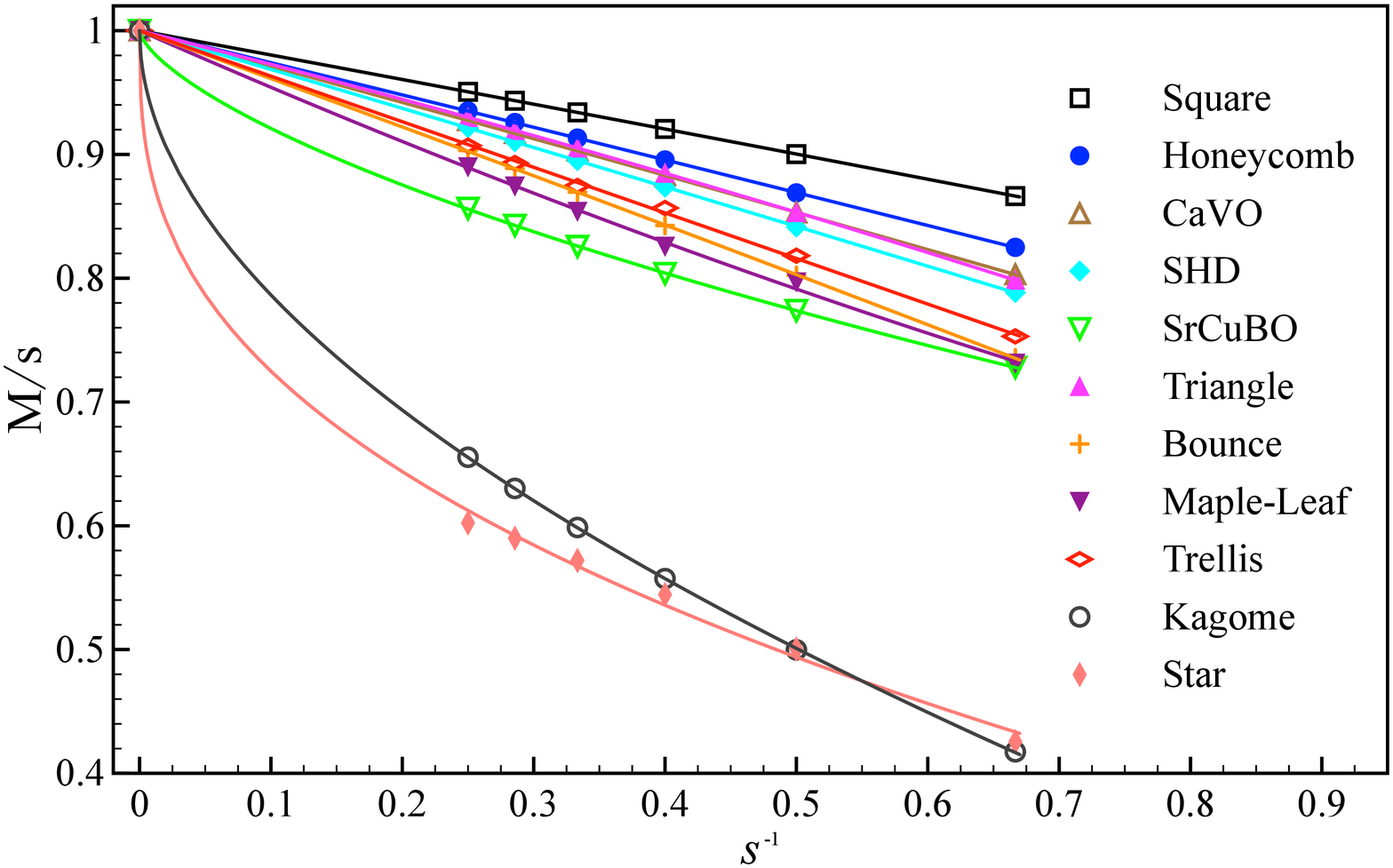}}
\caption{\label{fig3}(Color online) 
Results for the  order parameter, $M/s$, plotted as a function of $s^{-1}$  for $s \ge 3/2$. 
(Associated line fits to the data using the relations outlined in Table \ref{tab3} are shown also. 
For the sake of clarity, results for the order parameter obtained via extrapolation scheme I of 
Eq. (\ref{eqn_extra_2}) are included here only for the kagome and star lattices for the
$\sqrt{3}\times\sqrt{3}$ model state.)}
\end{center}
\end{figure}

\section{Conclusions}

We have explored the interplay between lattice topology and spin quantum number 
$s$ on the magnetic ordering of the HAFM on Archimedean lattices in this article. High-order
CCM calculations were carried out for all eleven Archimedean lattices by using highly 
intensive computational resources implemented on a parallel computing 
platform. It was shown in an appendix that the number of 
fundamental CCM clusters for the SUB$m$-$m$ approximation and the memory resources 
scale approximately exponentially with increasing level of approximation level, $m$.
Despite this, we have shown that the CCM can provide accurate results for the HAFM in 
all eleven Archimedean lattice, firstly by considering the limiting case of $s=1/2$.  
Excellent correspondence with those results of other approximate 
methods was observed for both the ground-state energy and order parameter for the
HAFM on all eleven Archimedean lattices, where they exist  for $s \le 4$. 
Accurate values for the ground-state energy are useful both experimentally
and theoretically, where for example they are useful in thermodynamic studies at finite
temperature \cite{bernu2001,Misguich2005,schmidt2017}.
See also  Ref. \cite{harada2011} for a QMC
treatment of the thermodynamic properties of the spin-one Heisenberg model on the square lattice
and Ref. \cite{johnston2011} for higher values of the spin quantum number.  Ref. 
\cite{mueller2018} details calculations using the Green function technique for the 
thermodynamics of the kagome-lattice Heisenberg antiferromagnet with arbitrary 
spin quantum number, and CCM results for the ground-state energy and the order parameter 
are more accurate than results presented in this paper. 
The ground-state energy and the order parameter are two fundamental parameters (among others) that 
define the behavior of quantum magnetic systems. Our results in Tables \ref{tab1} and \ref{tab2} 
therefore constitute an essentially quantitative reference for the Archimedean-lattice HAFMs with 
spin quantum number, $s \le 4$.

Scaling relations for the ground-state energy and order parameter as a function of $s$ 
were also presented for all of the Archimedean-lattice HAFMs. The scaling behavior for 
the ground-state energy was found to be the same as SWT, namely, that $E_g/(NJs^2)$ 
scaled with $s^{-1}$ to first order.  The scaling behavior for the order parameter was also 
found to be the same as that expected from linear SWT for most of the lattices, namely, that $M/s$ 
scaled with $s^{-1}$ to first order. 
Importantly, we remark again that self-consistent spin-wave theory calculations \cite{chubukov1992} 
showed that $M/s$ scales with $s^{-2/3}$ for the kagome lattice HAFM, whereas previous CCM 
results \cite{goetze2011} (found again here also) suggested that $M/s$ scales with $s^{-1/2}$. We
find also here that $M/s$ scales with $s^{-1/3}$ on the star lattice and with $s^{-2/3}$ 
on the  SrCuBO lattice. It would be interesting to see if self-consistent spin-wave 
theory might also detect this anomalous behavior for the star and SrCuBO lattices also,
which has not been noticed before for these two lattices, as far as we are aware. 

High-order SUB$m$-$m$ calculations for large values of $s$ suffer the double jeopardy of 
high computational demand and slow convergence with $m$. Thus, we assume the classical result $M/s 
\rightarrow 1$ in the limit $s \rightarrow \infty$ and we use CCM results for $s \le 4$ only
when evaluating the scaling behaviour. This is a limitation of these analyses, although it is 
particularly important only for the most extreme case of the star lattice. 
Despite this problem, expansion coefficients for the ground-state and order parameter 
with $s$ presented in Table \ref{tab3} were found to compare well to first order to 
the results of other approximate methods, for those cases where they are known. Furthermore, 
CCM results for $M/s$ for the kagome and star lattices for the $q=0$ model state were found 
to scale with $s^{-2/3}$. These results give us confidence that anomalous scaling behavior to first 
order for the order parameter does indeed occur for the kagome, star, and SrCuBO lattices. 
The SrCUBO-lattice HAFM is a special case ($J_2=1$) of the Shastry-Sutherland antiferromagnet 
\cite{shastry1981} that contains
antiferromagnetic nearest-neighbor bonds $J_1$ on the square lattice and with one
antiferromagnetic diagonal bond $J_2$ in each second square. 
Ref. \cite{ccm_high_order_SrCuBO_2005} states that the orthogonal-dimer state becomes 
the ground state at $J_2/J_1=1.477$ for the spin-half system. 
(Series expansions \cite{weihong1999} place this point at $J_2/J_1=1.447$
and recent results of iPEPS \cite{corboz2013} place it at $J_2/J_1=1.481$.)
Furthermore, N\'eel order was found to disappear in the range $J_2/J_1 = 1.14$ to 
$J_2/J_1 = 1.39$ for this spin-half system in Ref.  \cite{ccm_high_order_SrCuBO_2005},
where it was also posited that an intermediate regime may exhibit plaquette or columnar 
dimer ordering. (Results of iPEPS \cite{corboz2013} place this boundary at $J_2/J_1=1.307$.) 
The spin-half SrCuBO system might therefore  be 
``close'' (in some sense) to a disordered regime. However, it is still interesting and unexpected 
that the order parameter  demonstrates such anomalous scaling with $s$ for the
SrCuBO-lattice HAFM. Both the star and kagome lattices have an infinite number of ground states 
classically, and so unusual behavior is perhaps less unexpected. 


\section*{Acknowledgement}

We thank Prof. Johannes Richter for his insightful and interesting discussions relating to this work.

\appendix

\section{Computational Resources For High-Order CCM}

We now wish to consider the computational resources that 
are necessary in order to carry out high-order CCM calculations for the 
Archimedean lattices for all values of the spin quantum number, $s$. Note that 
$N_t$ is a measure of the number of terms and / or the memory usage 
of the CCM ket-state equation system; memory usage is a limiting 
factor of carrying out such large-scale CCM calculations.  
An additional complication for the bounce, trellis, and maple-leaf lattices 
is that we must carry out a dynamic search for the angle that yields the lowest 
ground-state energy. Thus for example, SUB8-8 for $s < 3$ and SUB7-7 for 
$s \ge 3$ was achievable only for the trellis lattice. 
The highest SUB$m$-$m$ level of approximation achieved by high-order CCM 
is shown in Table \ref{appendix_tab} for all of the Archimedean lattices and for 
$s \in \{1/2, 1, 3/2, \cdots ,4 \}$. 
The maple-leaf and bounce lattices both have large unit cells that contain
six sites and so the number of fundamental clusters $N_f$ increases strongly with 
increasing SUB$m$-$m$ approximation level. 
Generally the computational effort [$N_f$, $N_t$] increases with 
the spin quantum number $s$ and this is also shown in Table \ref{appendix_tab}.
For example, it was found previously \cite{bishop2016} that  
$N_f$ = 103097 for $s = 1/2$ at the SUB12-12 level of approximation on 
the honeycomb lattice, and $N_f$ = 538570 for $s = 9/2$ at the SUB10-10 
level of approximation for this lattice. Results for $N_f$ and $N_t$ (measured in 
terms of memory usage) on the square 
and kagome lattices and for $s = 1/2$ and $s=4$ are plotted as a function of 
SUB$m$-$m$ approximation level in Fig. \ref{fig_rqz}. We see that the number 
of fundamental configurations and the memory usage both grow approximately 
exponentially with increasing level of SUB$m$-$m$ approximation level (for a specific 
value of $s$). 
However, Table \ref{appendix_tab} also shows that the number of fundamental configurations
(and so also computer memory usage) saturates with increasing values of $s$ 
for the SUB$m$-$m$ scheme for a specific value of  $m$. (Note that $N_t$
refers to the number of terms contributing to the CCM equations rather than
memory usage in this table, although clearly the two measures are linked.) 
The limiting factor here is the amount of  computer resources available, thereby 
constraining the maximum value of $m$ that is possible. Furthermore, 
the SUB$m$-$m$ approximation constrains both the spatial separation of
sites on the lattice (i.e., $m$ contiguous sites) and also the maximum number 
of spin-flips (i.e., no more than $m$ spin-flips). However, the maximum
number of spin-flips per site is given by 2 for $s=1$ systems generally, whereas 
the maximum number of spin-flips per site is given by 8 for $s=4$ generally. 
It is the constraint on the maximum number of spin-flips for the SUB$m$-$m$
approximation that leads to ``saturation'' with increasing $s$ (keeping $m$
constant). However, we remark that the 
SUB7-7 approximation (and certainly higher orders of approximation) ought 
to provide reasonable results for $s \lesssim 4$. By contrast, the LSUB$n$ 
approximation does not restrict the number of spin-flips in any way and so we would 
not expect to observe a similar ``saturation'' with increasing $s$ for this approximation 
scheme (keeping $n$ constant). However, the LSUB$n$ scheme is far more intensive in terms of 
computer resources than the SUB$m$-$m$ scheme if we set 
$n=m$. Lower orders of LSUB$n$ approximation scheme are possible only 
compared to the SUB$m$-$m$ scheme, especially for larger values of the spin quantum number. 
Note also that higher orders of approximation can be reached for some cases 
only via highly intensive computational  methods that have been implemented for 
the CCM code using MPI. This approach ``shares'' out the cost (in terms of both 
the CPU time and memory usage) of finding and solving the CCM equations 
across all of processors that are used in parallel.


\begin{sidewaystable}
\caption{The value of $m$ for the highest level of SUB$m$-$m$ approximation achieved by high-order CCM for the HAFM on all 
Archimedean lattices and with $s \in \{1/2, 1, 3/2, \cdots ,4 \}$. The number of fundamental
configurations and the number of terms in the ket-state equations at this highest level of SUB$m$-$m$ approximation
are shown also in square brackets (i.e., [$N_f,N_t$]). ($N_t$ is presented in units of $10^{10}$ terms contributing to the
CCM equations here.)}
{\footnotesize
\begin{center}

\begin{tabular}{|l|c|c|c|c|c|c|c|c|c|}  \hline
            		&$~s=1/2~$ &$~s=1~$  &$~s=3/2~$ &$~s=2~$ &$~s=5/2~$ &$~s=3~$ &$~s=7/2~$ &$~s=4~$         \\ \hline
Square    &12 [766220, 1.89]    &10 [1056286, 0.91]    &10 [2006598, 1.55]    &10 [2231647, 1.62]    &10 [2253696, 1.66]    &10 [2253696, 1.66]    &9 [100011, 0.02]    &9 [100011, 0.02] \\ \hline
Honeycomb    &12 [103097, 0.10]    &10 [219521, 0.11]    &10 [461115, 0.20]    &10 [530418, 0.21]    &10 [538570, 0.22]    &10 [538570, 0.22]    &10 [538570, 0.22]    &10 [538570, 0.22] \\ \hline
CaVO    &12 [339887, 0.27]    &10 [806870, 0.37]    &10 [1744396, 0.72]    &10 [2026614, 0.77]    &10 [2061718, 0.79]    &10 [2061718, 0.79]    &10 [2061718, 0.79]    &9 [126642, 0.79] \\ \hline
SHD        &12 [241056, 0.18]    &10 [607383, 0.28]    &9 [93606, 0.01]    &9 [98778, 0.01]    &9 [98778, 0.01]    &9 [98778, 0.01]    &9 [98778, 0.01]    &9 [98778, 0.01] \\ \hline
SrCuBO    &10 [497361, 0.32]    &8 [253957, 0.05]    &8 [385889, 0.07]    &8 [403292, 0.07]    &8 [403292, 0.07]    &8 [403292, 0.07]    &8 [403292, 0.07]    &8 [403292, 0.07] \\ \hline 
Triangle    &10 [1054841, 12.19]    &8 [422399, 0.96]    &8 [661675, 1.41]    &8 [716945, 1.47]    &8 [725541, 1.49]    &8 [726350, 1.49]    &8 [726386, 1.49]    &8 [726387, 1.49] \\ \hline
Bounce   &8 [182422, 0.14]    &8 [1084987, 1.21]    &8 [1850973, 1.94]    &8 [2060029, 2.06]    &8 [2099003, 2.1]    &8 [2103493, 2.1]    &8 [2103719, 2.11]    &8 [2103721, 2.11] \\ \hline
Maple-Leaf    &8 [110145, 0.11]    &8 [1904330, 3.16]    &8 [3158718, 4.90]    &8 [3482740, 5.17]    &7 [371833, 0.21]    &7 [372037, 0.21]    &7 [372039, 0.21]    &7 [372039, 0.21] \\ \hline
Trellis    &8 [80293, 0.10]    &8 [1350826, 2.72]    &8 [2189409, 4.10]    &8 [2395909, 4.3]    &8 [2430164, 4.36]    &7 [242735, 0.17]    &7 [242736, 0.17]    &7 [242736, 0.17] \\ \hline
Kagome    &10 [238010, 0.85]    &8 [199717, 0.24]    &8 [358075, 0.40]    &8 [405130, 0.43]    &8 [414668, 0.44]    &8 [415871, 0.44]    &8 [415934, 0.44]    &8 [415935, 0.44] \\ \hline
Star    &10 [28897, 0.03]    &8 [45247, 0.02]    &8 [89469, 0.05]    &8 [105504, 0.05]    &8 [109592, 0.05]    &8 [110310, 0.05]    &8 [110373, 0.05]    &8 [110374, 0.05] \\ \hline
 \hline
\end{tabular}
\end{center}
}
\label{appendix_tab}
\end{sidewaystable}

\begin{figure}
\begin{center}
\epsfxsize=13cm
\centerline{\epsffile{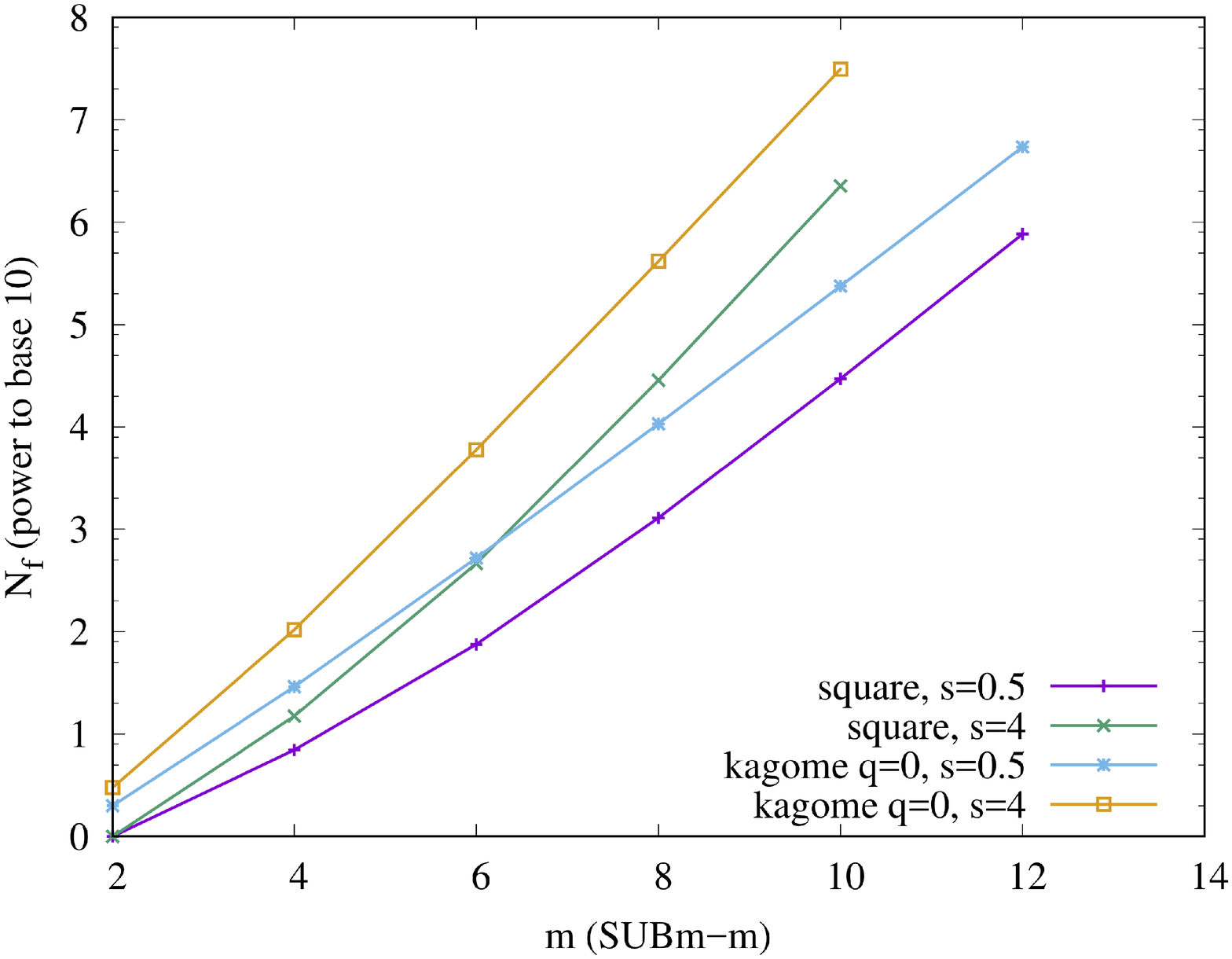}}
\centerline{\epsffile{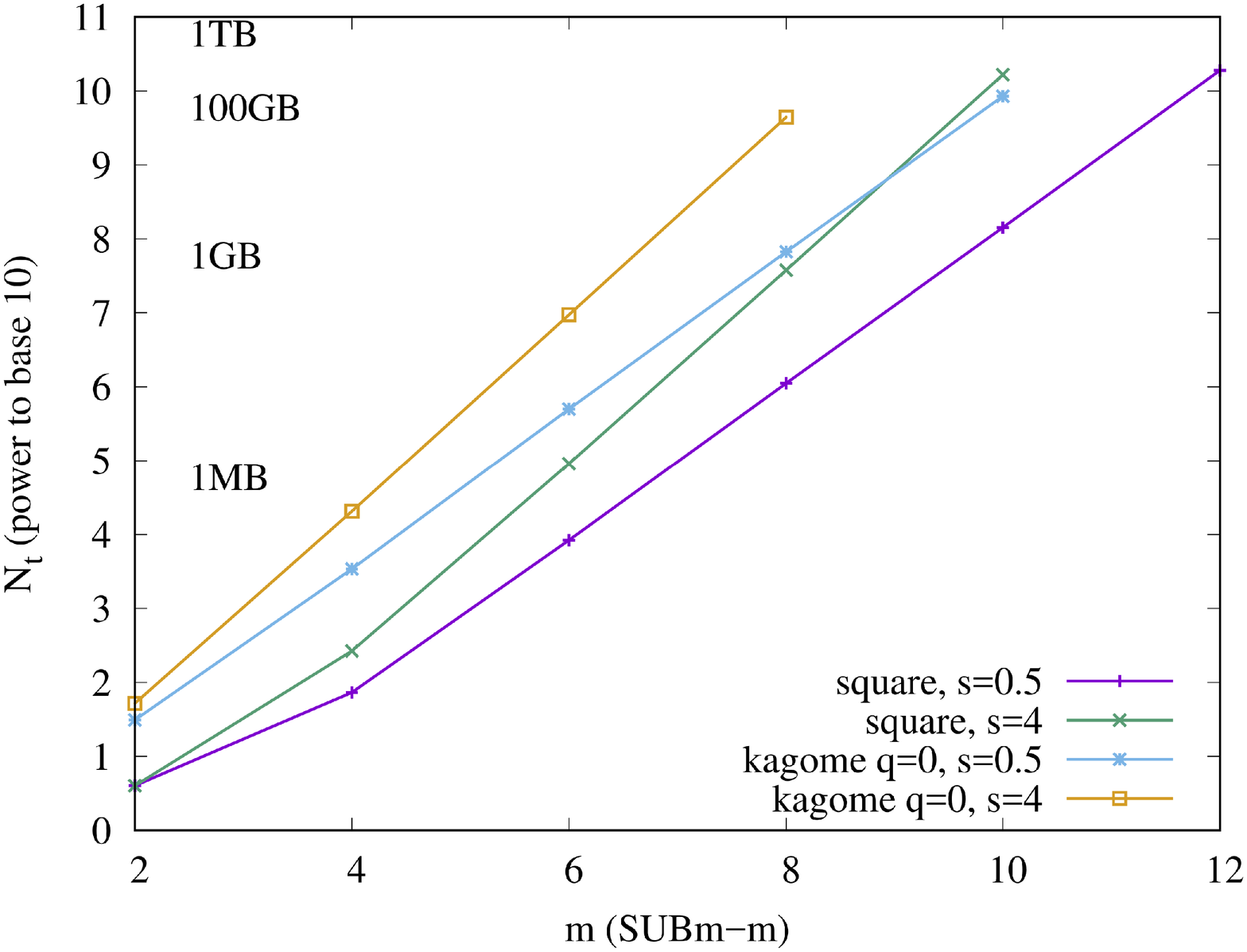}}
\caption{\label{fig_rqz}(Color online) Illustration of the number of fundamental configurations $N_f$ (top) and the 
number of terms $N_t$ (measured in terms of memory usage here) of the ket equation system (bottom) 
plotted as a function of the approximation $m$. The square and the kagome lattices with $s=\frac{1}{2}$ 
and  $s=4$  have been chosen as examples.  Note that the $N_f$ and  the $N_t$ data are scaled 
logarithmically (base 10) and that  the data for even numbers of $m$ are shown only.}
\end{center}
\end{figure}

\pagebreak

\end{document}